# Engineering of electronic and magnetic modulations in gradient functional oxide heterostructures

L. Schüler[1], Y. Sievert[1], V. Roddatis[2], U. Ross[3], V. Moshnyaga[1], and F. Lyzwa[4,5]

[1]I. Physikalisches Institut, Georg-August-Universität Göttingen, Friedrich-Hund-Platz 1, 37077 Göttingen, Germany

[2]GFZ Helmholtz Centre for Geosciences, Telegrafenberg, 14473 Potsdam, Germany

[3] IV. Physikalisches Institut, Georg-August-Universität-Göttingen, Friedrich-Hund-Platz 1, 37077 Göttingen, Germany

[4]Department of Physics, Photon Factory, University of Auckland, 38 Princes Street, Auckland 1010, New Zealand

[5]Te Whai Ao Dodd-Walls Centre for Photonic and Quantum Technologies, New Zealand

Advanced interface engineering provides a way to control the ground state of correlated oxide heterostructures, which enables the shaping of future electronic and magnetic nanodevices with enhanced performance. An especially promising and rather new avenue is to find and explore low-dimensional phases of structural, ferroic and superconducting origin. In this multimodal study, we present a novel dynamic growth control method that enables synthesizing compositionally graded superlattices (SLs) of $(LaMnO_3)_{10}/(SrMnO_3)_{10}$ (LMO/SMO), in which the layers gradually change their composition between LMO and SMO with gradient ***G*** values ranging from 0 to 100%. This leads to strong modulations in the material's electronic properties and of the two-phase ferromagnetic (FM) behavior. In particular, we observe that ***G*** suprisingly has almost no impact on the emergent high-temperature FM phase; in contrast, the low-temperature volume-like FM phase increases drastically with higher G-factors and thus can serve as a precise marker for chemical composition on a nanoscale. Focusing on the interfacial charge transfer found at sharp SMO/LMO interfaces (***G***=0), we observe that for higher ***G***-factors a long-range charge modulation develops, which is accompanied by an insulator-to-metal transition. These findings showcase ***G*** as a crucial control parameter that can shape the superlattice's intrinsic properties and provide a perspective for designing functional oxide heterostructures with artificially disordered interfaces.

**Introduction**

Heterostructures of strongly correlated oxides reveal a variety of novel electronic and magnetic orders that are absent in the constituent layers, possess great application potential for the next-generation electronic and spintronic devices [1], [2], [3], [4]. Moreover, these systems serve as a material platform for exploring tunable Rashba physics [5], understanding the mechanism of high-$T_c$ superconductivity in e.g. cuprate/manganite multilayers [6], [7], controlling insulator-metal/superconductor transitions [8] and phononic bandgap engineering [9], [10], and exploring emergent magnetism in, e.g., $(LaMnO_3)_m/(SrMnO_3)_n$ ($LMO_m/SMO_n$) superlattices (SLs) [11], [12].

Developing thin film methods to control emerging electronic and magnetic orders in oxide heterostructures is essential for better understanding the underlying physics and making these structures more attractive for technological applications [13], [14]. One prominent way is the application of epitaxial strain, governed by the choice of the substrate. For instance, it has been reported that compressive (tensile) strain suppresses (promotes) the ferromagnetic order in LMO/SMO structures [15]. Moreover, biaxially strained two-dimensional (2D) Ruddlesden-Popper heterostructures of $Sr_{n+1}Ti_nO_{3n+1}$ reveal a ferroelectric instability [16] not observed in bulk materials. Furthermore, substrate termination can govern the appearance of 2D metallic interfaces [17]. Another control parameter is the thickness ratio of the constituent layers in $A_m/B_n$ SLs, e.g. a high thickness ratio "m/n" leads to an enhancement of the charge transfer (CT) in $LMO_m/(SrTiO_3)_n$ [18] and $LMO_m/SMO_n$ SLs [11]. In the latter case, the enhanced CT from the electron-rich LMO to the electron-poor SMO layers promotes the emerging high-$T_C$ ferromagnetic (HTP) phase, located at the SMO/LMO interface.

Yet another approach to control the macroscopic properties of heterostructures is to modify the interfaces between adjacent layers through chemical substitution/intermixing, as successfully demonstrated in semiconducting heterostructures [19], [20], [21]. Namely, such compositionally graded interface regions were shown to be useful for electronic bandgap engineering in AlAs/GaAs [19], for thermal conductivity control in Si/Ge [20] as well as for controlling the charge flow in halide perovskite photovoltaic cells [21]. Gradient/graded heterostructures that possess strain gradients can result in strong flexoelectric fields [22];

recently, a gradient-strain-induced ferroelectric state at room temperature has been reported to arise in La$_2$NiMnO$_6$/La$_2$CoMnO$_6$ ferromagnetic SLs [23], which is especially interesting for multiferroic systems.

Interface engineering has long been a subject of extensive research within the complex oxide community. However, the study of gradient SLs, i.e. periodic structures with compositionally graded layers has been far less investigated for oxide heterostructures, despite its magnetic and electronic control potentials. A limiting factor in the fabrication of gradient structures is the ability to monitor and control the film growth process, thus deriving *in situ* information on the structure and electronic properties of films during the growth. Typical monitoring techniques include 'reflection high-energy electron diffraction' (RHEED), 'grazing-incidence small-angle x-ray scattering' (GISAXS) [24], 'reflectance-difference spectroscopy' (RDS), p-polarized reflectance spectroscopy and spectroscopic ellipsometry (SE) [25], [26]. However, the information obtained primarily addressed the structural properties of the layers, rather than their electronic properties. Recently, we demonstrated the capabilities of optical ellipsometry as an in-situ monitoring and control method for the growth of complex oxides heterostructures. A relation between the detected optical signal and the structural, magnetic and electronic properties of the thin film heterostructures could be established [11], [12], [27] with atomic layer precision [28].

In this paper, we report an advanced, *dynamic* growth control method (*dyna-MAD* [29]) to design gradient (or compositionally graded) interfaces in LMO/SMO SLs, resulting in a modulation of the charge distribution and emergent magnetic phases within the crystal. The degrees of gradient have been varied at a spatial scale D=2-10 u.c. from G=0% (i.e. atomically sharp interfaces in between the LMO$_{10}$/SMO$_{10}$ layers) to G=100% (i.e. fully La/Sr intermixed interfaces and LSMO$_{10}$/SLMO$_{10}$ layers). The impact of gradients on the two-phase magnetic behavior has been studied with a special focus on the interplay of atomic intermixing and interfacial charge transfer. It is demonstrated that optical ellipsometry is a valuable *in situ* diagnostic for gradient perovskite oxide SLs.

**Results and discussion**

The general concept of gradient LaMnO$_3$/SrMnO$_3$ (LMO/SMO) superlattices (SLs) with varying amount of gradient/graded layers is displayed in Fig. 1. A nominal **G**=0 gradient represents heterostructures exhibiting sharp interfaces with estimated interface roughness of ~1 u.c, as previously reported [11], [12].

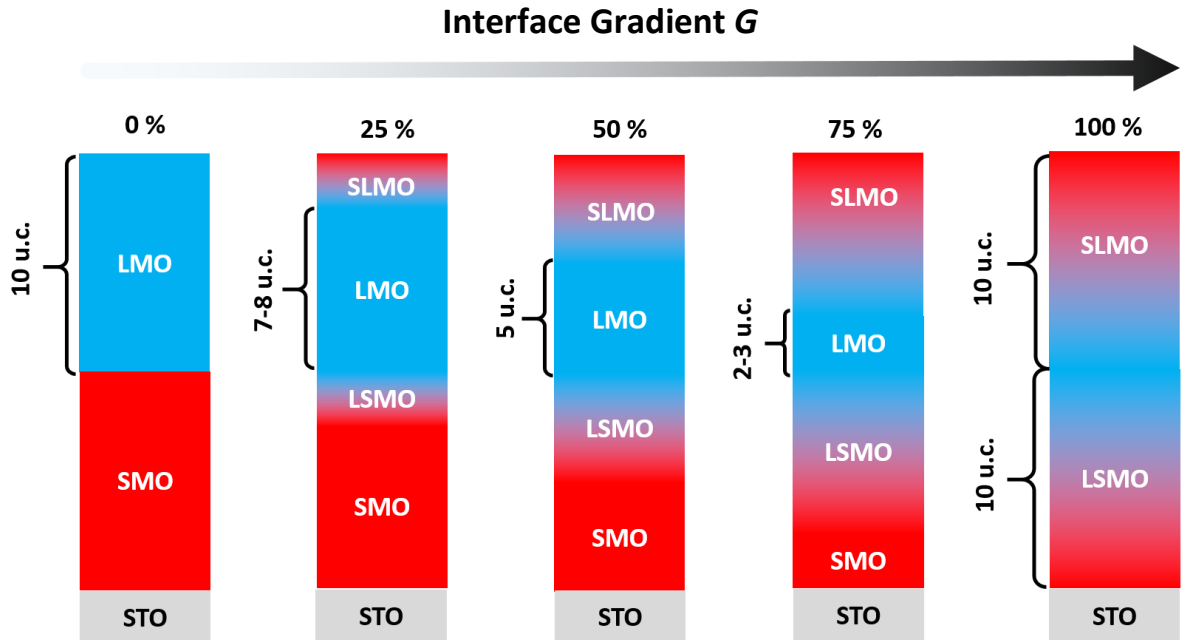

**Figure 1.** Principle and nominal structure of a bilayer in graded LMO/SMO SLs with interfaces consisting of graded La-doped SMO (**SL**MO) and graded Sr-doped LMO (**LS**MO). All LMO/SMO SLs are grown on SrTiO$_3$ (100) substrates (shown in gray color).

Note that a nominal **G**=0 gradient does not exclude interfacial intermixing, as indicated by the XRR results (see below). The reason is that XRR as well as TEM have been applied to real samples with micro (substrate steps) and macro (sample dimensions for XRR are 5x10 mm$^2$) thickness/composition modulations. They are additionally influenced for TEM samples by the beam spread over the sample thickness in cross-section. These material issues can be accounted for in simulations of La/Sr XRR profiles along the growth direction, resulting in effective gradient values of **G**~10-20 %, or in "graded" interface thicknesses of D$_{SLMO, LSMO}$=1-2 u.c. for SLs with unintentionally graded interfaces.

In this study, we focused on the impact of increasingly intermixed interfaces on the charge transfer as well as the magnetic and structural properties of SLs. The bilayer thickness for all SLs was set at 20 u.c., consisting of 10 u.c. of LMO and 10 u.c. of SMO for an SL with nominally $G$=0% interface smearing (see Fig. 1). Hence, the SL with 50 % gradient (GL50) consists of "pure" (LMO)$_5$ and (SMO)$_5$ blocks, separated by 5 u.c. thick blocks where the Sr-doping in the LMO gradually changes from x=0 to x=1, i.e. starting with pure LMO and ending with pure SMO; we named these blocks (**LS**MO)$_5$. When the 5 u.c.-thick graded region develops from the pure (SMO)$_5$ block towards the pure (LMO)$_5$, these blocks can be viewed as La-doped SMO or (**SL**MO)$_5$. Thus, the nominal overall composition of the SL with N=7 repetitions and 50 % gradient interfaces is: [(LMO)$_5$/(LSMO)$_5$/(SMO)$_5$/(SLMO)$_5$]$_N$. For the SLs with 25 %, 75 % and 100 % gradient amount, the thickness of the LSMO and SLMO gradient interfaces formally consists of 2.5, 7.5 and 10 u.c, respectively, as depicted in Fig. 1.

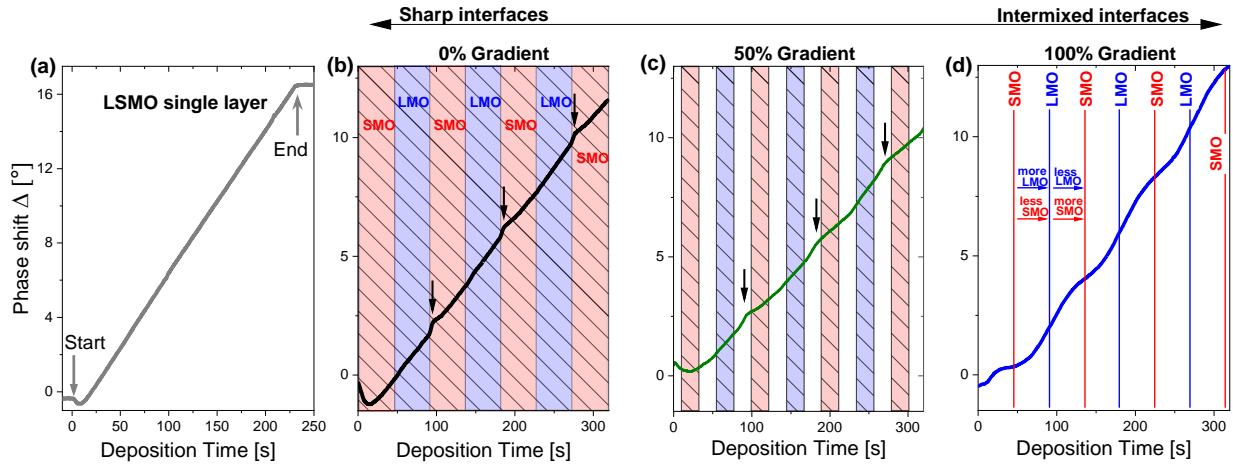

**Figure 2. (a)** Ellipsometric phase shift angle, Δ, as a function of deposition time and layer thickness, as detected during the growth for a single La$_{0.7}$Sr$_{0.3}$MnO$_3$/SrTiO$_3$ (100) (LSMO/STO) film with a thickness of 5 nm (13 u.c.). The first stage growth signal of ca. 3. u.c. is typical for MAD-technique (more details in Refs. [12], [11]); then, a linear increase of the film thickness at a constant deposition rate of ca. 0.08 nm/s is observed. **(b-d)** The ellipsometry signal for an LaMnO$_3$/SrMnO$_3$ (LMO/SMO) superlattice on STO with sharp interfaces (0% gradient) is shown in (b), 50% gradient in (c) and 100% gradient in (d). The black arrows highlight the increase in the ellipsometry signal that result from the interfacial charge transfer.

Figure 2 presents the concrete examples of the MAD-grown LSMO film and the SLs with graded interfaces grown by dyna-MAD and monitored by *in situ* optical ellipsometry; further details on this method can be found in Ref. [29] and Fig. SM-1 of the supplementary materials [30]. The time dependencies of the ellipsometric phase shift, Δ(*t*), are shown for an 18.5 nm thick

LSMO single film (Fig. 2a)) as a reference, for an SL with **G**=0 % intermixing (i.e., not intentionally mixed interfaces) dubbed as GL0 (Fig. 2 b), for G=50 % as GL50 (Fig. 2c), and for **G**=100 % (i.e., fully intermixed gradient regions) GL100 in Fig. 2d). During the initial growth step of approximately 10-13 s, a dip in Δ(*t*) is observed for the LSMO film and GL0 and GL50 samples. This feature is typical for thin films grown by means of MAD-technique on STO substrates (see e.g. [11]) and is caused by the initial formation of small 2D islands, resulting in diffuse light scattering during ellipsometry [31]. After overcoming this transition region with an effective thickness of $D_0$~3 u.c., the LSMO film exhibits linear Δ(t) and Δ(D) dependencies due to a constant growth rate and a constant refractive index, as detailed in [12], [28].

For the case of consecutively deposited LMO and SMO layers one over the other (Fig. 2b), a similar constant increase in Δ(t) or Δ(D) is expected, here with two different slopes dD/dt for the LMO and SMO layers due to their different refractive indices. Indeed, the LMO layers in Fig. 2 b) show a linear increase in the phase shift with time and thickness, as expected. When starting to grow the SMO layer on top, however, a modified optical Δ(t, D) response has been observed. Namely, a sudden increase in Δ(*t*) (indicated by arrows in Fig. 2b and 2c) is seen within the first 2 u.c. of the SMO layer close to the preceding LMO layer at the SMO(top)/LMO(bottom) interface. The LMO(top)/SMO(bottom) interface shows no change in the optical signal. This electronic modification, assigned to the interfacial charge transfer (CT) from the electron-rich LMO to the electron-poor SMO layer, leads to an emergent high-$T_C$ ferromagnetic phase in these LMO/SMO SLs [12], [11]. For the progressively intermixed **G**=50% interfaces shown in Fig. 2c), this well-defined CT-bump, localized within the gradient region, is getting suppressed. Finally, for **G**=100% the CT-bump vanishes and the ellipsometry signal exhibits a wavy shape (see Fig. 2d)) with the maximum slope close to the "pure" LMO and a minimum slope close to the "pure" SMO.

The local and global structure of the SLs has been also characterized ex situ by high-resolution HAADF-STEM and XRD/XRR; the results are presented in Fig. 3 and Fig. 4, respectively. The TEM images in Fig. 3 clearly display the differences between the GL0 (Fig. 3a)) and graded GL50 (Fig. 3a)) and GL100 (Fig. 3b)) samples. Namely, the GL0 shows well-defined LMO and SMO layers with the thickness vayring in the range 10±1 u.c. along the growth direction; within the zoomed region in Fig. 3 a) the GL0 looks as $LMO_{11}/SMO_9$.

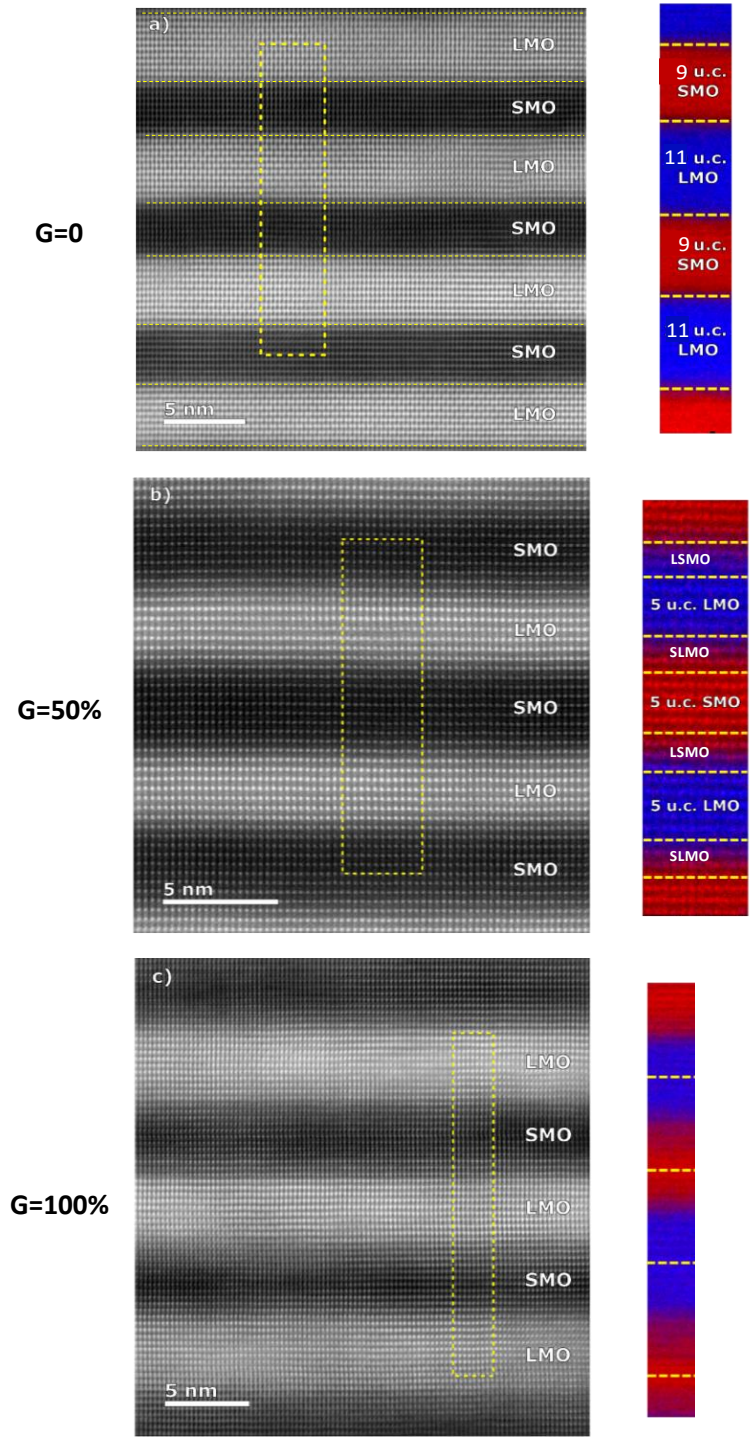

**Figure 3.** HAADF-STEM images of representative SLs with nominally 0% **(a)**, 50% **(b)** and 100% **(c)** gradients. The corresponding right panels in **(a)**, **(b)** and **(c)** represent EELS maps of the yellow-marked regions in the HAADF-STEM images. **(a)** SL0 exhibits LMO and SMO layers with thickness 11 and 9 u.c., respectively, as well as sharp interfaces with RMS~1 u.c. yielding the above mentioned gradient value G=10-20%. **(b)** SL50 shows LMO and SMO layers with thickness 5 u.c., separated by graded regions comprising LSMO and SLMO layers with thickness 3-4 u.c. **(c)** SL100 reveals strongly La/Sr-intermixed regions with thickness about 10 u.c.; the position of "nominally pure" LMO and SMO layers is marked by the corresponding lines.

Sharp interfaces with a root-mean-square (RMS) roughness of ~1 u.c. can be recognized, thus, contributing to the above assumed gradient value **G**=10-20% in real SLs. The GL50 sample clearly exhibits increasingly intermixed interface regions in HAADF-STEM additionally intensified by atomically resolved EELS color maps with La atoms marked in "blue" and those of Sr in "red". According to the EELS maps in Fig. 3 the GL50 consists of 5 u.c. "pure" LMO and SMO layers, separated by 3 u.c. of LSMO and 4 u.c. of SLMO layers. The actual gradient amount in this nominally 50% gradient SL is estimated as a ratio between the sum of gradient regions, i.e. of LSMO and SLMO, and the whole bilayer thickness consists of G=($D_{LSMO}$+$D_{SLMO}$) / ($D_{LMO}$+$D_{LSMO}$+$D_{SMO}$+$D_{SLMO}$)=7 u.c./(16-17) u.c.= 41-44 %. This indicates that, although there is an uncertainty in the bilayer thickness of 3-4 u.c. (the nominal bilayer has to be 20 u.c. thick), the actual gradient amount is close to the nominal 50 %. As for the GL100 sample (Fig. 3 c)), one can see an apparently poor contrast between LMO and SMO layers; the ADF-STEM image is dominated by the presence of LSMO and SLMO "gray" regions.

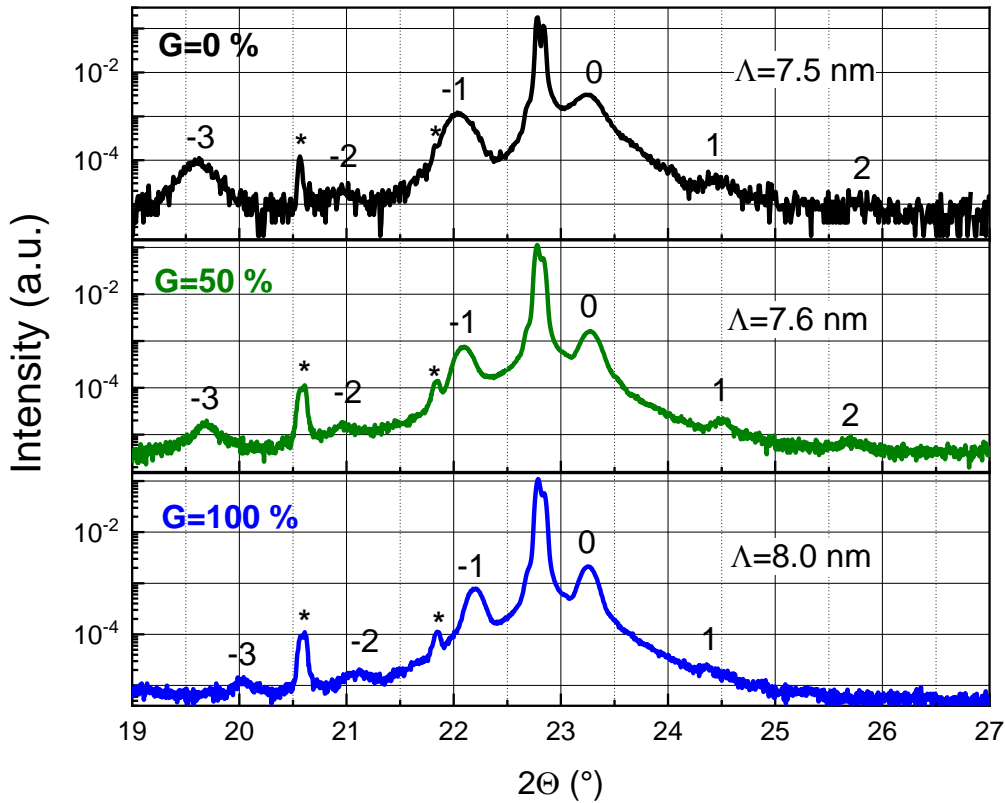

**Figure 4.** XRD patterns normalized to the STO(200) peak of SLs with G=0, 50 and 100 % measured around the STO(100) peak. Λ represents the calculated superlattice period or bilayer thickness.

The corresponding EELS map shows a rather continuous La/Sr modulation with a "nominally pure" LMO and SMO layers marked by lines in Fig. 3 c). A more detailed EELS spectra analysis can be found in the supplementary materials [30], Fig. SM-2.

The XRD patterns (Figure 4) indicate epitaxially strained heterostructures for the different gradient values (0-100 %), which exhibit superlattice peaks with diffraction order n=0, ±1, ±2, ±3, with positions being close to each other for different SLs. The superlattice period $\Lambda$ (or bilayer thickness) as calculated according to the superlattice formula $2\Lambda*(\sin\theta_n - \sin\theta_0)=\pm n\lambda$ [32], varies in the range of $\Lambda$=7.5-8 nm (see Fig. 4), which indicates a deviation in the bilayer thickness of about 1 u.c. for all samples. These values are very close to those directly observed by TEM (see Fig. 3) as well as those determined by XRR simulations (see Fig. 5 below), demonstrating a good reproducibility of the dyna-MAD growth of SLs.

Figure 5 presents the XRR data of representative SLs with 0%, 50% and 100 % gradients, along with the corresponding fitting curves (shown in pink) obtained using the program GenX [33]. This simulation accounts for the optical properties of the constituting materials, i.e. LMO and SMO, based on their differing electron densities. Moreover, the most crucial fitting parameters used are the layer thickness, interface and surface roughness. In Fig. 5, one can clearly see superlattice modulations for all samples, from positions of which the bilayer thicknesses of $\Lambda$=7.3-8 nm were evaluated, in good agreement with the XRD data (see Fig. 4). Moreover, the LMO/SMO interface roughness, obtained from the simulations and which also accounts for the doping gradient at the interfaces, was found to increase with higher G-value. Finally, the simulated surface RMS roughness of SLs was determined to be in the range of RMS=0.4-0.8 nm, being in good agreement with RMS-values determined from AFM measurements ( [30], Fig. SM-3).

The XRR simulations allow us to model the LMO/SMO electronic profiles along the growth direction, as exemplified in the insets in Fig. 5 for GL0, GL50, and GL100 . Importantly, according to GenX simulations of GL0 and GL50, the presence of LMO and SMO layers is indicated by periodic oscillations of scattering length density (SLD) between the maximum value, $SLD_{max}$=1.755, and the minimum value, $SLD_{min}$=1.506, which correspond to the theoretical values of pure LMO and SMO, marked in blue and red, respectively.

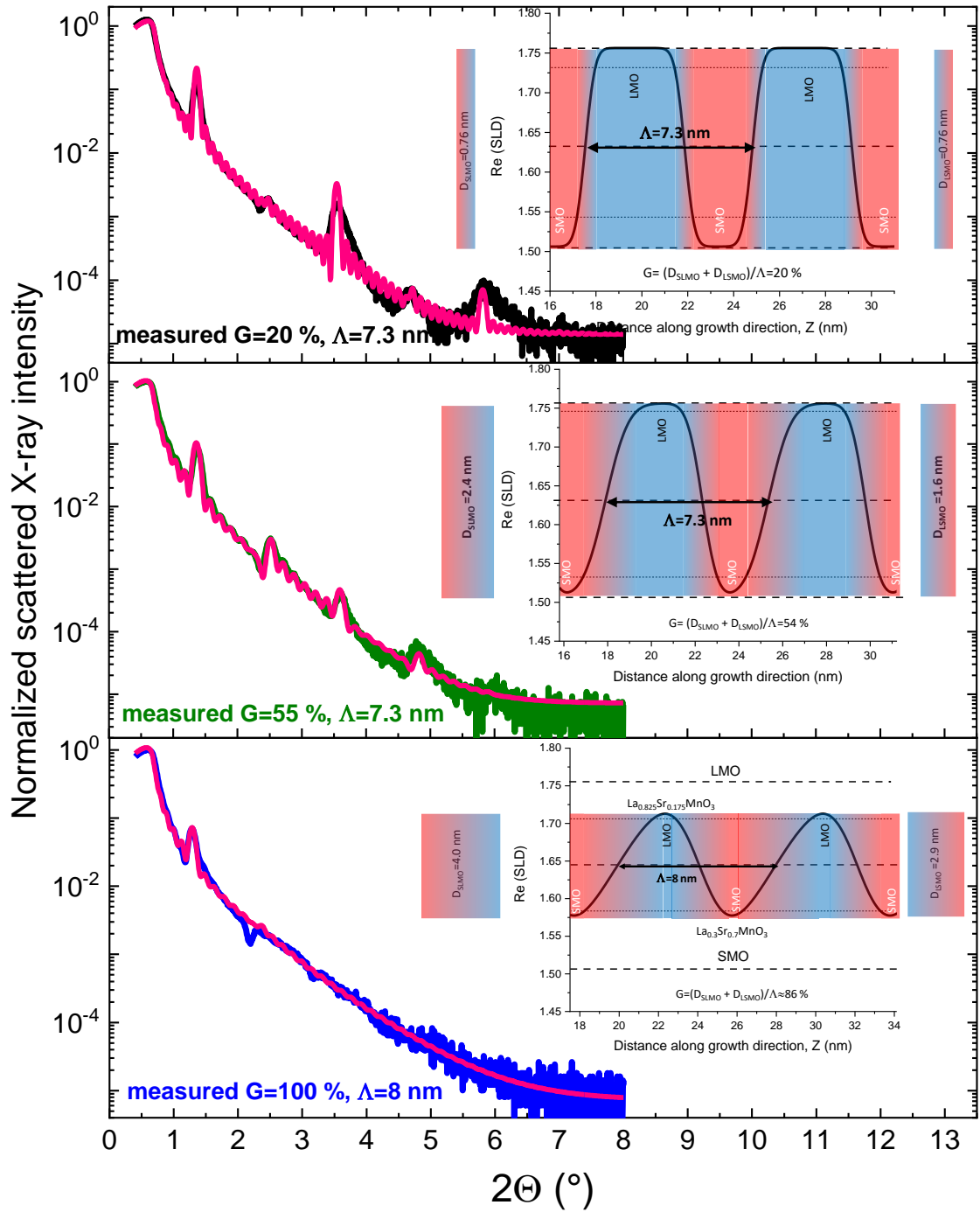

**Figure 5.** Small-angle X-ray reflection (XRR) on SLs with **G**=0, 50 and 100 % gradient values and the reconstructed LMO/SMO profiles (insets). The fits are shown in pink color.

The average electron density in such SLs is close to that in a $La_{0.5}Sr_{0.5}MnO_3$ film, which is in accordance with the designed equal thicknesses of the LMO and SMO layers of 10 u.c. each.

In between the pure LMO and SMO, one finds gradient regions where the electron density (represented by the color in Fig. 5) continuously changes from "red" to "blue", corresponding to the intermixed **SL**MO layers, and from "blue" to "red", depicting the graded **LS**MO layers. The thickness of such graded interfaces in the GL0 sample was estimated to be about $D_{SLMO}=D_{LSMO}\approx0.7$-$0.8$ nm~2 u.c., thus corresponding to an interface roughness of ≈1 u.c. of each LMO(top)/SMO(bottom) and SMO(top)/LMO(bottom) interface. The overall estimated gradient amount for the GL0 sample is $G_0=100\%*(D_{LSMO}+D_{SLMO})/\Lambda\approx1.5$ nm/$7.3$ nm$\approx20\%$. Note, that according to this definition of the G-value, there cannot be any SL with G=0 due to the presence of small but still finite interface roughness, which, even in the best samples cannot be < 1 u.c..

For GL50, enlarged **SL**MO and **LS**MO gradient regions with thicknesses $D_{SLMO}=2.4$ nm and $D_{LSMO}=1.6$ nm are observed. The amount of gradient in the GL50 sample of $G=100\%*(D_{LSMO}+D_{SLMO})/\Lambda=4$ nm/$7.3$ nm$=54\%$, as determined by XRR, agrees well with the designed value of G=50 %. In the fully intermixed G100 sample, the electron density oscillates not in-between the LMO and SMO, but rather the SLD oscillations occur between the values corresponding to the compositions of $La_{0.83}Sr_{0.17}MnO_3$ and $La_{0.3}Sr_{0.7}MnO_3$. This means, that there seem to be no pure LMO and SMO layers in SLs: the former contains about 17 % of Sr and the latter about 30 % of La. Thus, the average composition of the GL100 sample is shifted from the half-doped LSMO to that of $La_{0.57}Sr_{0.43}MnO_3$. This is unlikely to be solely accounted to imperfections in the LMO and SMO control of feeding rates, but is related to the more efficient growth of LMO compared to that of SMO. The amount of gradient in the nominal **G**=100% sample can be only roughly estimated as $G=100\%*(D_{LSMO}+D_{SLMO})/\Lambda\approx100\%*(6.9$ nm/$8$ nm$)\approx86\%$. This value, being close to the nominal **G**=100 %, can be nevertheless considered as a realistic one because of the large error in the determination of the LMO and SMO layer thickness. Note, that a not well-defined thickness of "pure" LMO and of "pure" SMO layers (see Fig. 5), being considerably thinner than the gradient regions, amounts to about 0.5-0.6 nm, which corresponds to ~1.5 u.c. only. Remarkably, in the graded SLs (GL50 and GL100) the interfaces look strongly asymmetric. For instance, as seen in Fig. 5, the LMO/SMO graded interface is thicker than that of SMO/LMO by about 0.8 nm in GL50 and by about 1.1 nm in GL100. Such asymmetry is probably related to a larger diffusion of smaller Sr cations into the growing LMO layer from the LMO/SMO interface as vice versa fewer diffusion of large La atoms into the SMO layer from the SMO/LMO interface.

Now we turn to the time/thickness dependence of the *in situ* measured optical ellipsometry signal provides information on the film growth and interfacial charge transfer in the LMO/SMO SLs. Optical ellipsometry, as any other infrared/optical spectroscopy method, is sensitive to the complex refractive index of a material, which in turn is related to its electronic properties and the charge density as well (for more details refer to supplementary materials of [11]). In Figure 1a), the linear $\Delta(t)$ time dependence in a single LSMO film (after the initial transition zone of ~2-3 u.c.) and constant phase shift rate $d\Delta/dt \approx 0.08°/sec = 0.38°/1$ u.c. reflect homogeneous electronic properties of the growing LSMO thin film, i.e. a constant charge density. For the LMO/SMO SLs with sharp interfaces (see top panel of Fig. 6), **G**=0%), a pronounced CT peak in the phase shift rate $d\Delta/dt$ dependence, marked by the arrow, is located within the SMO layer close to the SMO(top)/LMO(bottom) interface in good agreement with previous reports [12], [11]. This indicates an enhanced charge density within the first ~2 u.c. of SMO due to CT from the LMO layer. Remarkably, the same (or very similar) phase shift rate in the G0 sample as in the optimally doped homogeneous LSMO film, i.e. $d\Delta/dt \approx 0.08°/sec = 0.38°/1$ u.c., points out its LSMO-like origin. With increasing amount of gradient, the CT-peak broadens and its intensity is suppressed in the GL25 and GL50 samples. Nevertheless, this broadened CT-peak is fully located in the gradient regions, which indicates that the CT itself occurs within the gradient regions.

For GL100, as one can see in Fig. 6 (bottom panel), a distinct CT-peak close to the SMO/LMO interface is no more observable and the presence of CT in the fully intermixed sample is questionable. However, GL100 also shows a clear modulation of the phase shift rate and, thus, of the charge density in between the minimal value $d\Delta/dt \approx 0.025°/sec = 0.1°/1$ u.c., which is close to the phase shift rate of SMO (ref. 12), and maximal value of $d\Delta/dt \approx 0.08°/sec = 0.38°/1$ u.c. chracteristic for optimally doped LSMO. This allows us to suggest that the CT in the fully intermixed SL is not limited to the interface but rather spreads throughout the entire SL period according to the evaluated La/Sr-profile in Fig. 5 c). Moreover, with increasing gradient, the density of the transferred charge, being spread out within the whole SL, seems to be reduced compared to the interface CT localized close to the SMO/LMO interface.

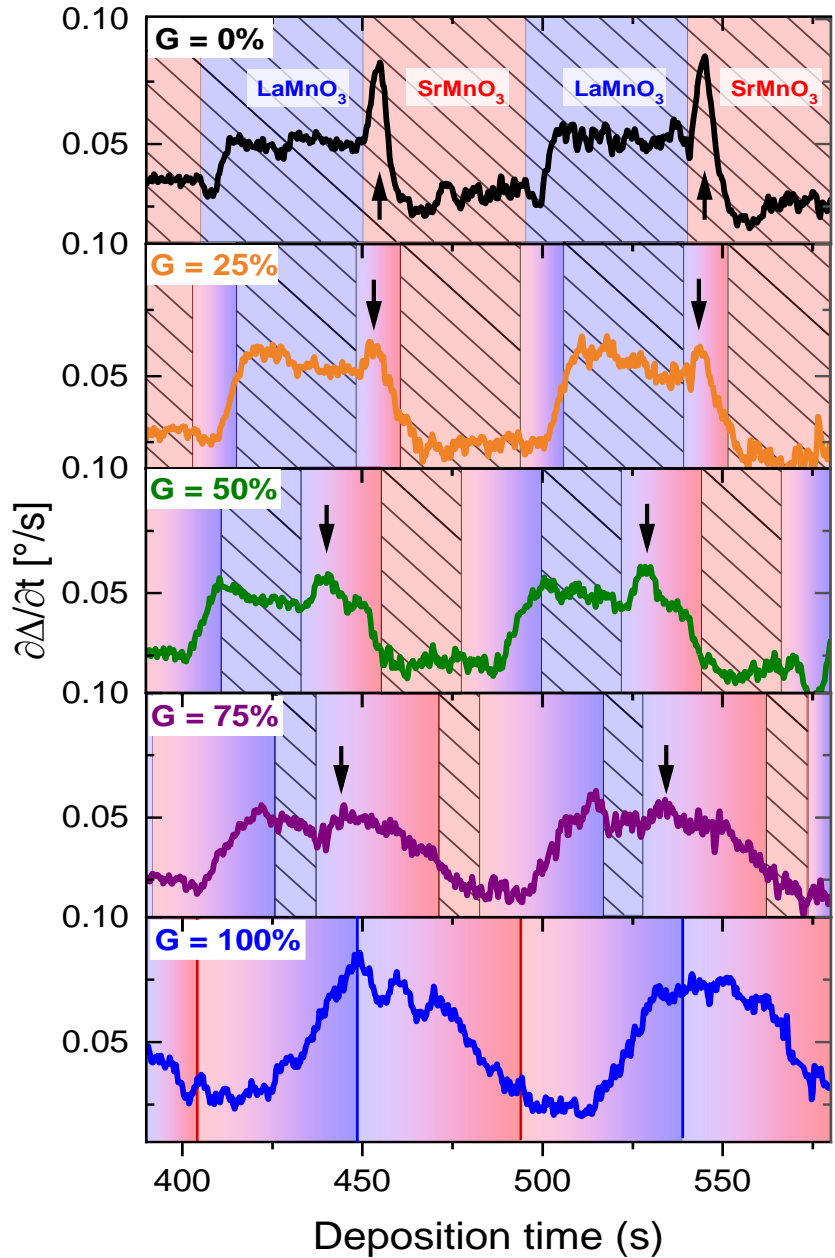

**Figure 6**. Time/thickness dependence of the phase shift rate dΔ/dt for LMO/SMO heterostructures with different nominal values of gradients **G**=0, 25, 50, 75 and 100 %. The pure LMO and SMO regions are indicated by the hatched regions, and the CT-peak by the arrows.

Note, that it is not trivial to clearly distinguish between pure electronic and chemical (Sr-doping) contributions to the CT; the latter also leads to a modification of the charge density. Our optical ellipsometry method probes at $\lambda$=0.63 μm (excitation wavelength of the He-Ne laser), which is sensitive to metallic or insulating behavior via spectral weight transfer mechanism [34], independent whether it was caused by adding an additional charge or by chemical doping.

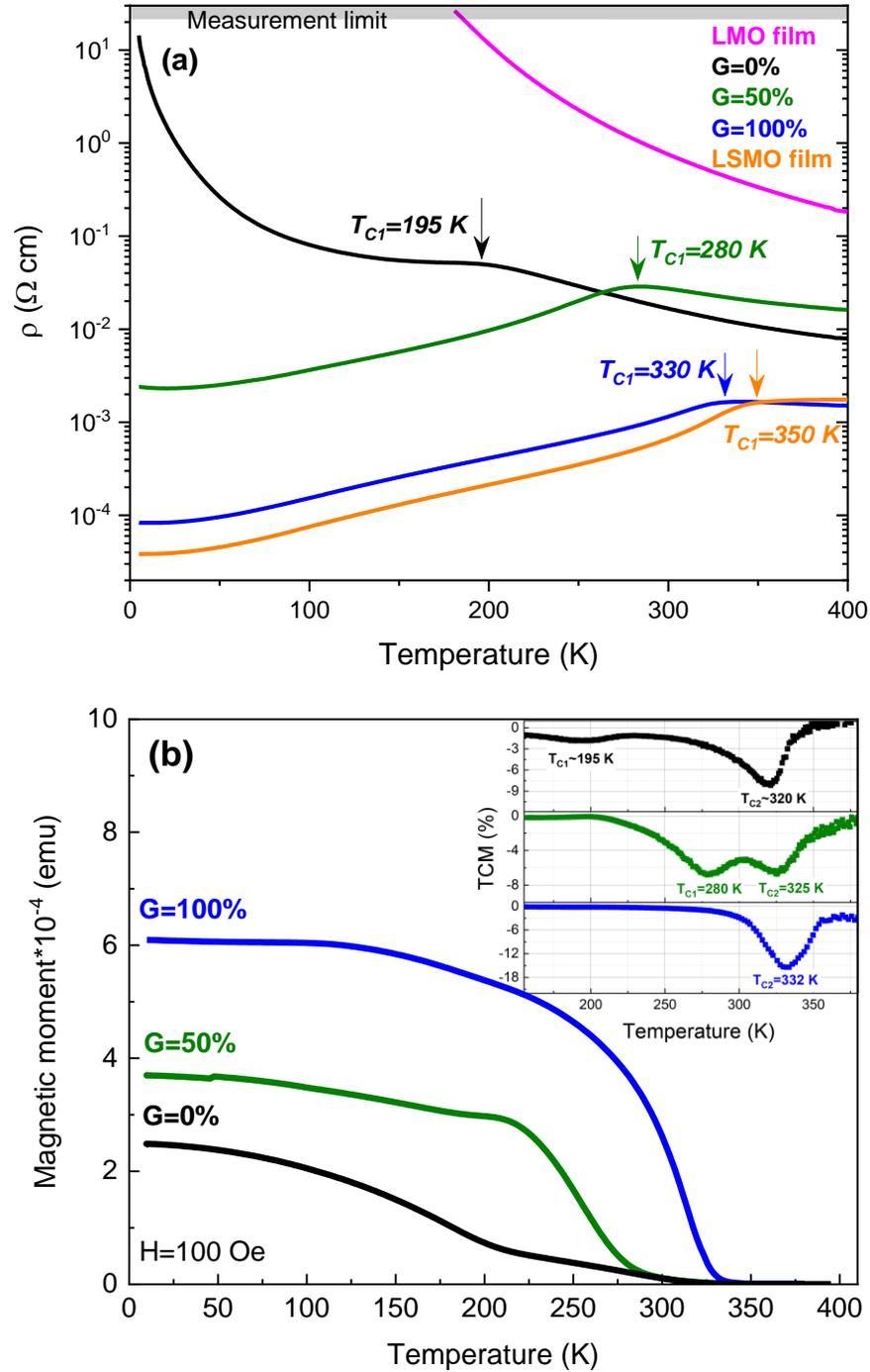

**Figure 7. (a)** Electrical resistance and **(b)** field-cooled magnetic moment versus temperature in gradient LMO/SMO superlattices. The inset Iin (b) displays the TCM-function, from the minimum of which $T_C$ values were determined. Shown are the the nominal values for **G**.

Figure 7 shows the results of the electrical resistivity (a) and magnetic moment (b) measurements. One can clearly see that the GL0 and GL100 samples reveal insulating and metallic behavior, respectively, in close agreement with insulating undoped LaMnO$_3$ (LMO) and metallic optimally doped LSMO films also shown in Fig. 7 (a) for comparison. The GL50 sample reveals an insulator-to-metal transition at 280 K, however, the residual resistance at 5 K is relatively large and dR/dT<0, as expected for an insulator. All this suggests the presence of a metal-to-insulator transition in gradient SLs for G>50 %, as will be further described below. A complete set of electrical resistance data for all gradient superlattices can be found in Fig. SM-4 [30]. In Fig 7 (b), the magnetic moment of SLs with **G**=0, 50 and 100 % is plotted as a function of temperature. For the GL0 and GL50 samples, one can see two well-separated phase transitions with $T_{C1}$ < $T_{C2}$, that denote the low-$T_{C1}$ (LTP) and high-$T_{C2}$ (HTP) FM phases, respectively. The Curie temperatures were determined from the minimum of the temperature coefficient of magnetization TCM= 100%*(1/M(T)*(dM/dT), shown in the inset of Fig. 7. In contrast to the GL0 and GL50 samples, displaying a two-phase FM behavior, the fully intermixed GL100 reveals only one FM transition at $T_C$=332 K. Moreover, the **G**=100% sample exhibits the largest magnetic moment, which decreases with decreasing **G**. Thus, the peculiarity of the GL100 showing a more homogeneous charge distribution along the bilayer thickness (see Fig. 6) is additionally reflected by the exsistence of a single-phase ferromagnetic behavior. Interestingly, the magnetic Curie temperatures of the LTP ($T_{C1}$) match well the metal-to-insulator transition temperatures derived from electrical measurements as well as the bump in R(T) for GL0 sample, pointing out the dominating role of the LMO-like LTP phase in electron transport.

In Figure 8, we present the Curie temperatures (left scale) and electrical conductivity (right scale) for all studied SLs with different values of gradients, which highlights the main result of this article. The HTP shows a relatively weak dependence on the amount of gradient varying in the range $T_{C2}$=320-332 K. In contrast, the Curie temperature of the LTP phase reveals a pronounced nonlinear dependence on the intermixing of LMO and SMO at the interfaces: it increases from $T_{C1}$=193 K up to $T_{C1}$=332 K by increasing degrees of gradients from **G**=20 to 100 %. The $T_{C1}$(G) dependence extrapolated to **G**=0 gives $T_{C1}$(0)≈170 K, which fits nicely to the Curie temperature of a single 40 nm thick LMO/STO film [12], [11]. Thus, the LTP, corresponding in the clean limit

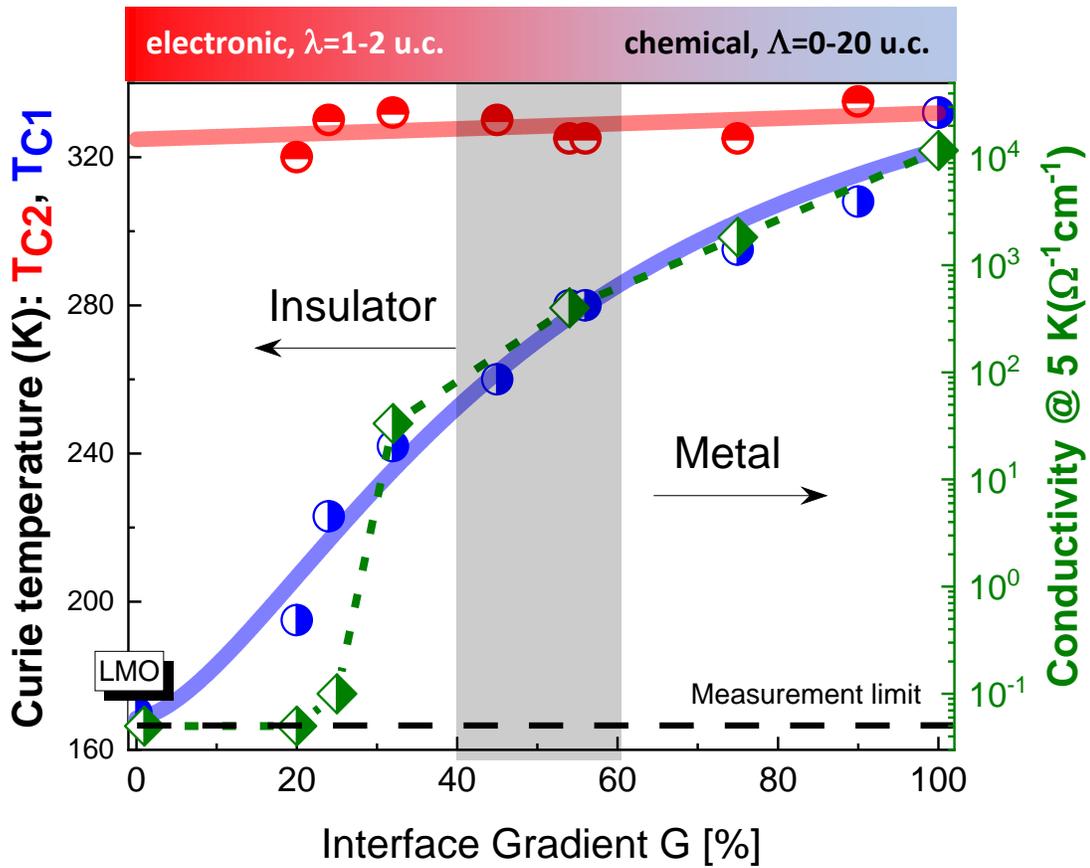

**Figure 8.** Curie temperatures $T_C$ (left scale) and electrical conductivity at 5 K (right scale) of LMO/SMO SLs with different degrees of gradient interfaces $G=100\%*(D_{SLMO}+D_{LSMO})/(D_{LMO}+D_{SMO})$. All data were taken from the temperature dependences of magnetic moment carried out at an external magnetic field H=100 Oe. The magnetic data points are fitted by a linear (for high-$T_C$) and a nonlinear logistic (low-$T_C$) approach. The low (high) values of conductivity at 5 K denote insulating (0<G<40 %) and metallic (60<G<100 %) ground states of SLs. The gray zone for 40<G<60 % mark the electronically inhomogeneous region.

G=0 to the undoped LMO layers, likely originates from the FM volume-like contribution of LMO in SLs. By progressively increased intermixing (0<G<100 %), the LTP increases due to Sr-doping of the LMO layers, and reaches the HTP value for **G**~80-100 % (i.e. fully intermixed LMO/SMO SL). Qualitatively similar is the dependence of low-temperature electrical conductivty $\sigma$(5 K) on the gradient value, **G**: the conductivity strongly increases with higher **G**, indicating the presence of an insulating phase for 0 < G < 40 % and a metallic phase for 60 % < G < 100 %. The ground state of SLs with intermediate values 40% < G < 60 % is poorly defined pointing out the metallic/insulating phase coexistence (see [35]).

In a previous report [12], a similar HTP and LTP behavior as a function of the interface density $1/\Lambda$ has been observed for digital $LMO_m/SMO_n$ SLs with LMO/SMO ratio m/n=1, 2 and different layer thicknesses n=1-6 u.c. Namely, the high-$T_{C2}$ was found to be almost independent on the interface density, but the low-$T_{C1}$ did depend on $1/\Lambda$, revealing a curve very similar to that shown in Fig. 8. Indeed, for high interface density or small bilayer thickness $\Lambda$=m+n (for n<3) both LTP and HTP phases merge together, yielding a single FM phase corresponding to the doped LSMO. For a small interface density or larger bilayer thickness (3<n<10), the low-$T_{C1}$ decreases with decreasing interface density and approaches the Curie temperature of LMO. Considering the fact that both interface density and gradient value are inversely proportional to the bilayer thickness $\Lambda$, the similarity between the LTP behavior shown in Fig. 8 and that found in Ref. [12] is not surprising. Under the assumption of a small but finite interface roughness, a gradient of **G**=2 u.c./6 u.c.≈33 % in an SL with very thin layers, e.g. $LMO_3/SMO_3$, can be obtained even for sharp LMO/SMO interfaces with a roughness ~1 u.c. According to Fig. 8, the LTP in such a hypothetical SL should possess $T_{C1}$≈245 K, which is comparable with $T_{C1}$≈260 K measured in the 3/3 SL (see Ref. [12]).

Finally we point out that optimally doped LSMO possesses the highest $T_C$=360 K [36]. Our graded LMO/SMO SLs show a high-$T_C$ ferromagnetic behaviour with $T_C$~320-330 K, whereas the interfacial emerging phase with $T_C$~350 K [11], [12] originates from atomically sharp SMO/LMO interfaces. We thus propose the formation of an interfacial layer with magnetic properties similar to the optimally doped LSMO, provided by an electron transfer from LMO to SMO. For SLs with **G**=100%, there are no real interfaces and the La/Sr intermixing dominates the electronic distribution, yielding an LSMO-like magnetic behavior.

**Conclusions**

$LaMnO_3/SrMnO_3$ superlattices with gradually changing La/Sr interfaces were synthesized by using a novel dynamic growth method (dyna-MAD). This is, to the best of our knowledge, the first detailed study demonstrating a gradual compositional change of an ordered to an increasingly disordered struture in an complex oxide system. We have investigated their structural, optical, magnetic and electronic properties by means of a series of in-situ and ex-situ characterization techniques. The new control parameter **G** (i.e, the gradient value) is introduced as the ratio between the thickness of LSMO and SLMO intermixed layers to the total bilayer thickness and characterized by X-ray reflectivity measurements and simulations. A two-phase magnetic behavior with high-$T_{C2}$ (HTP) and low-$T_{C1}$ (LTP) ferromagnetic phases was observed, with HTP weakly dependent on G, whereas LTP strongly increases with increasing G and merging with HTP for **G**~100 %. This LTP can thus serve as a precise marker of chemical composition in such heterostructures. In addition, a metal-to-insulator transition at around **G** ~40-60% has been observed. The role and relationship between interfacial charge transfer and chemical intermixing (La/Sr) are discussed, both exhibiting comparable length scales.

Our current interpretation demonstrates that by gradually increasing the disorder in LMO/SMO heterostructures using dyna-MAD (which is not trivial via other deposition techniques), the emergent low-dimensional ferromagnetic phase can be investigated and controlled. This approach can be of further use to explore other low-dimensional phases of ferroic and superconducting origin, by precisely changing the desired doping. In return, this can have a tremendous effect on the electronic and phonoic bandstructure as well as on the heat capacity.

*Experimental Methods*

**Thin film growth**. SLs with a nominal composition of [(LMO$_3$)$_{10\ u.c.}$/(SMO$_3$)$_{10u.c.}$]$_7$ and graded interfaces have been grown on SrTiO$_3$(100) (STO) substrates via dyna-MAD [29], schematically shown in Fig. SM-1 [30]. Within a conventional MAD process [12] the metalorganic precursors (e.g., La- and Mn-acetylacetonates in case of LMO growth) are dissolved together in an organic solvent (e.g., N,N-Dimethylformamide); the solution is then sprayed onto a heated substrate ($T_{Sub}$=400-900 °C) usually by compressed air under atmospheric conditions. An oxide film (e.g. LMO) grows on the substrate as a result of a heterogeneous pyrolysis reaction. The deposition of various complex oxide films has been demonstrated, such as LMO, SMO and La$_{1-x}$Sr$_x$MnO$_3$ by using a mixture of precursors taken in a certain molar ratio and dissolved in the same solvent [31].

To grow the graded SLs by means of dyna-MAD, we implemented a simultaneous programmable dynamic control of two liquid channels, i.e. for LMO and SMO. This was realized by using a combination of a homebuilt 3D-printed precursor injection system and of the precursor injection system from a conventional MAD (SyrDos liquid dosing units by HiTec Zang GmbH). The precursors are thus not pre-mixed in a common precursor solution (as in conventional MAD) but, instead, they are dynamically mixed inside the deposition chamber. Following this procedure, a precise control and adjustment of the gradual transition between the LMO and SMO layers has been achieved by simultaneous change of the LMO and SMO precursor feeding rates with deposition time. For example, after growing a pure SMO layer with thickness $D_{SMO}$=5-10 u.c., the SMO feeding rate has been reduced with a simultaneously increasing LMO rate, resulting in a deposition of a mixed manganite, e.g. $Sr_{1-x}La_xMnO_3$. The doping level "x" becomes time- and, thus, thickness-dependent, changing from x=0 to x=1 continuously or in steps of $\Delta x$=0.1-0.2 on the time scale corresponding to the desired thickness of a mixed layer $D_{SLMO}$=2-10 u.c. Afterwards, a pure LMO layer with thickness $D_{LMO}$=5-10 u.c. has been grown from the LMO precursor channel and the bilayer of such a graded SL has been finalized by the graded $La_{1-x}Sr_xMnO_3$ layer with a thickness $D_{LSMO}$=2-10 u.c.

*In situ monitoring.* Thin film growth has been monitored by means of a home-built *in situ* optical ellipsometry in the "Polarizer-Modulator-Sample-Analyzer" configuration [37] using a He-Ne laser with a wavelength λ=632.8 nm. This approach has been proven to be a very sensitive probe of the optical/electronic properties of the film, yielding a sub-monolayer precision [28]. Further information can be found in [11], [12].

*Structural Characterization.* The crystallographic structure and layer thickness were studied *ex situ* by X-Ray diffraction and reflection (XRD and XRR) using a Bruker Advance diffractometer D8 with Cu-$K_{\alpha 1,\alpha 2}$ radiation. High-resolution Scanning Transmission Electron Microscopy (STEM) using a Thermo Fisher Titan G2 ETEM and probe-corrected Titan microscopes operating at 80-300 kV was performed on lamellae prepared by a lift-out method using a Thermofisher Scientific (TFS) Helios G4UC dual beam instrument operated at 30 kV, 16 kV, 5 kV and 2 kV, respectively. High Angle Annular Dark Field (HAADF) and ADF images in cross-section geometry were aquired from focused ion-beam prepared samples on a Thermo Fisher Titan G2 ETEM, as well as a probe-corrected Titan acquired using a TFS Themis. In addition to structural imaging, a spatially resolved chemical characterization was performed by recording electron energy-loss (STEM-EELS) spectra spanning a region containing the core-loss from Oxygen-K (532 eV) to well above the Strontium-L edge (1940 eV). Energy Loss Spectroscopy (EELS) and acquisition of Spectrum Images (SIs) were done using a Gatan Continuum 1063 spectrometer.

***Magnetization measurements*** have been carried out in the temperature range, $T$ = 5–400 K, for an external magnetic field, $H$ = 100 Oe, by using a SQUID magnetometer (MPMS XL, "Quantum design").

***Transport Measurements.*** The resistance has been measured via the four-point probe method in the temperature range $T$ = 5–400 K, using a Quantum Design Physical Properties Measurement System (PPMS).

***Atomic Force Microscopy images*** were taken using a Bruker Innova device.

**Acknowledgements**

The authors thank Yurii G. Pashkevich and J.P. Bange for fruitful discussions. This work was financially supported by the Deutsche Forschungsgemeinschaft (DFG, German Research Foundation) - 217133147/SFB 1073, project Z02. F.L. acknowledges support from the Swiss National Science Foundation through an Early.Postdoc.Mobility Fellowship with Project No. P2FRP2-199598. The authors thank the European Regional Development Fund and the State of Brandenburg for the Themis Z TEM (part of Potsdam Imaging and Spectral Analysis (PISA) Facility).

# References

[1] H. Y. Hwang, Y. Iwasa, M. Kawasaki, B. Keimer, N. Nagaosa and Y. Tokura, "Emergent phenomena at oxide interfaces," *Nature Materials,* vol. 11, p. 103–113, February 2012.

[2] F. Hellman, A. Hoffmann, Y. Tserkovnyak, G. S. D. Beach, E. E. Fullerton, C. Leighton, A. H. MacDonald, D. C. Ralph, D. A. Arena, H. A. Dürr, P. Fischer, J. Grollier, J. P. Heremans, T. Jungwirth, A. V. Kimel, B. Koopmans, I. N. Krivorotov, S. J. May, A. K. Petford-Long, J. M. Rondinelli, N. Samarth, I. K. Schuller, A. N. Slavin, M. D. Stiles, O. Tchernyshyov, A. Thiaville and B. L. Zink, "Interface-induced phenomena in magnetism," *Rev. Mod. Phys.,* vol. 89, no. 2, p. 025006, June 2017.

[3] J. Chakhalian, J. W. Freeland, A. J. Millis, C. Panagopoulos and J. M. Rondinelli, "Colloquium: Emergent properties in plane view: Strong correlations at oxide interfaces," *Rev. Mod. Phys.,* vol. 86, no. 4, p. 1189–1202, October 2014.

[4] C. A. F. Vaz, F. J. Walker, C. H. Ahn and S. Ismail-Beigi, "Intrinsic interfacial phenomena in manganite heterostructures," *Journal of Physics: Condensed Matter,* vol. 27, p. 123001, February 2015.

[5] P. Noël, F. Trier, L. M. Vicente Arche, J. Bréhin, D. C. Vaz, V. Garcia, S. Fusil, A. Barthélémy, L. Vila, M. Bibes and J.-P. Attané, "Non-volatile electric control of spin-charge conversion in a SrTiO3 Rashba system," *Nature,* vol. 580, p. 483–486, 2020.

[6] N. Driza, S. Blanco-Canosa, M. Bakr, S. Soltan, M. Khalid, L. Mustafa, K. Kawashima, G. Christiani, H.-U. Habermeier, G. Khaliullin, C. Ulrich, M. Le Tacon and B. Keimer, "Long-range transfer of electron-phonon coupling in oxide superlattices," *Nature Materials,* vol. 11, p. 675–681, 2012.

[7] E. Perret, C. Monney, S. Johnston, J. Khmaladze, F. Lyzwa, R. Gaina, M. Dantz, J. Pelliciari, C. Piamonteze, B. P. P. Mallett, M. Minola, B. Keimer, T. Schmitt and C. Bernhard, "Coupled Cu and Mn charge and orbital orders in YBa2Cu3O7/Nd0.65(Ca1-ySry)0.35MnO3 multilayers," *Communications Physics,* vol. 1, p. 45, August 2018.

[8] A. D. Caviglia, S. Gariglio, N. Reyren, D. Jaccard, T. Schneider, M. Gabay, S. Thiel, G. Hammerl, J. Mannhart and J.-M. Triscone, "Electric field control of the LaAlO3/SrTiO3 interface ground state," *Nature,* vol. 456, p. 624–627, December 2008.

[9] F. Lyzwa, A. Chan, J. Khmaladze, K. Fürsich, B. Keimer, C. Bernhard, M. Minola and B. P. P. Mallett, "Backfolded acoustic phonons as ultrasonic probes in metal-oxide superlattices," *Phys. Rev. Materials,* vol. 4, no. 4, p. 043606, April 2020.

[10] S. G. Jeong, A. Seo and W. S. Choi, "Atomistic Engineering of Phonons in Functional Oxide Heterostructures," *Advanced Science,* vol. 9, p. 2103403, 2022.

[11] J. P. Bange, V. Roddatis, L. Schüler, F. Lyzwa, M. Keunecke, S. Lopatin, V. Bruchmann-Bamberg and V. Moshnyaga, "Charge Transfer Control of Emergent Magnetism at SrMnO3/LaMnO3 Interfaces," *Advanced Materials Interfaces,* vol. 9, p. 2201282, 2022.

[12] M. Keunecke, F. Lyzwa, D. Schwarzbach, V. Roddatis, N. Gauquelin, K. Müller-Caspary, J. Verbeeck, S. J. Callori, F. Klose, M. Jungbauer and V. Moshnyaga, "High-Tc Interfacial Ferromagnetism in SrMnO3/LaMnO3 Superlattices," *Advanced Functional Materials,* vol. 30, p. 1808270, 2020.

[13] D. G. Schlom, L.-Q. Chen, X. Pan, A. Schmehl and M. A. Zurbuchen, "A Thin Film Approach to Engineering Functionality into Oxides," *Journal of the American Ceramic Society,* vol. 91, pp. 2429-2454, 2008.

[14] J. Mannhart and D. G. Schlom, "Oxide Interfaces—An Opportunity for Electronics," *Science,* vol. 327, pp. 1607-1611, 2010.

[15] H. Yamada, P.-H. Xiang and A. Sawa, "Phase evolution and critical behavior in strain-tuned ${\text{LaMnO}}_{3}{\text{-SrMnO}}_{3}$ superlattices," *Phys. Rev. B,* vol. 81, no. 1, p. 014410, January 2010.

[16] C.-H. Lee, N. D. Orloff, T. Birol, Y. Zhu, V. Goian, E. Rocas, R. Haislmaier, E. Vlahos, J. A. Mundy, L. F. Kourkoutis, Y. Nie, M. D. Biegalski, J. Zhang, M. Bernhagen, N. A. Benedek, Y. Kim, J. D. Brock, R. Uecker, X. X. Xi, V. Gopalan, D. Nuzhnyy, S. Kamba, D. A. Muller, I. Takeuchi, J. C. Booth, C. J. Fennie and D. G. Schlom, "Exploiting dimensionality and defect mitigation to create tunable microwave dielectrics," *Nature,* vol. 502, p. 532–536, 2013.

[17] N. Nakagawa, H. Y. Hwang and D. A. Muller, "Why some interfaces cannot be sharp," *Nature Materials,* vol. 5, p. 204–209, 2006.

[18] J. Garcia-Barriocanal, F. Y. Bruno, A. Rivera-Calzada, Z. Sefrioui, N. M. Nemes, M. Garcia-Hernández, J. Rubio-Zuazo, G. R. Castro, M. Varela, S. J. Pennycook, C. Leon and J. Santamaria, ""Charge Leakage" at LaMnO3/SrTiO3 Interfaces," *Advanced Materials,* vol. 22, pp. 627-632, 2010.

[19] H. Xu, X. Zhou, G. Xu, Q. Du, E. Wang, D. Wang, L. Zhang and C. Chen, "Photoreflectance study of AlAs/GaAs gradient period superlattice," *Applied Physics Letters,* vol. 61, pp. 2193-2195, November 1992.

[20] Y. Liu, J. Hao, A. Chernatynskiy, G. Ren and J. Zhang, "Effect of period length distribution on the thermal conductivity of Si/Ge superlattice," *International Journal of Thermal Sciences,* vol. 170, p. 107157, 2021.

[21] H. W. Qiao, S. Yang, Y. Wang, X. Chen, T. Y. Wen, L. J. Tang, Q. Cheng, Y. Hou, H. Zhao and H. G. Yang, "A Gradient Heterostructure Based on Tolerance Factor in High-Performance Perovskite Solar Cells with 0.84 Fill Factor," *Advanced Materials,* vol. 31, p. 1804217, 2019.

[22] P. Zubko, G. Catalan and A. K. Tagantsev, "Flexoelectric Effect in Solids," *Annual Review of Materials Research,* vol. 43, pp. 387-421, 2013.

[23] Y. Jiang, X. Wu, J. Niu, Y. Zhou, N. Jiang, F. Guo, B. Yang and S. Zhao, "Gradient Strain-Induced Room-Temperature Ferroelectricity in Magnetic Double-Perovskite Superlattices," *Small Methods,* vol. 7, p. 2201246, 2023.

[24] G. Renaud, R. Lazzari, C. Revenant, A. Barbier, M. Noblet, O. Ulrich, F. Leroy, J. Jupille, Y. Borensztein, C. R. Henry, J.-P. Deville, F. Scheurer, J. Mane-Mane and O. Fruchart, "Real-Time Monitoring of Growing Nanoparticles," *Science,* vol. 300, pp. 1416-1419, 2003.

[25] I. P. Herman, "Optical Diagnostics for Thin Film Processing," *Annual Review of Physical Chemistry,* vol. 54, pp. 277-305, 2003.

[26] R. W. Collins, J. S. Burnham, S. Kim, J. Koh, Y. Lu and C. R. Wronski, "Insights into deposition processes for amorphous semiconductor materials and devices from real time spectroscopic ellipsometry," *Journal of Non-Crystalline Solids,* Vols. 198-200, pp. 981-986, 1996.

[27] P. Ksoll, R. Mandal, C. Meyer, L. Schüler, V. Roddatis and V. Moshnyaga, "Emergent double perovskite phase at $\mathrm{LaMnO}_3/\mathrm{LaNiO}_3$ interfaces: Coupled charge transfer and structural reconstruction," *Phys. Rev. B,* vol. 103, no. 19, p. 195120, May 2021.

[28] F. Lyzwa, P. Marsik, V. Roddatis, C. Bernhard, M. Jungbauer and V. Moshnyaga, "In situ monitoring of atomic layer epitaxy via optical ellipsometry," *Journal of Physics D: Applied Physics,* vol. 51, p. 125306, March 2018.

[29] F. Lyzwa, L. Schüler and V. Moshnyaga, "Dynamic growth control of complex oxide heterostructures," *Submitted,* 2024.

[30] L. Schüler, Y. Sievert, V. Roddatis, U. Ross, V. Moshnyaga and F. Lyzwa, "supplementary materials".

[31] M. Jungbauer, "Atomic Layer Design of Electronic Correlations in Perovskite Heterostructures," 2016.

[32] D. K. Bowen and B. K. & Tanner, "High Resolution X-Ray Diffractometry And Topography," *CRC Press.,* vol. 1st. Edition, pp. 141-142, February 1998.

[33] A. Glavic and M. Björck, " 3: the latest generation of an established tool," *Journal of Applied Crystallography,* vol. 55, p. 1063–1071, August 2022.

[34] A. I. Lobad, R. D. Averitt, C. Kwon and A. J. Taylor, "Spin–lattice interaction in colossal magnetoresistance manganites," *Applied Physics Letters,* vol. 77, pp. 4025-4027, December 2000.

[35] M. Uehara, S. Mori, C. H. Chen and S.-W. Cheong, "Percolative phase separation underlies colossal magnetoresistance in mixed-valent manganites," *Nature,* vol. 399, p. 560–563, 1999.

[36] A. Urushibara, Y. Moritomo, T. Arima, A. Asamitsu, G. Kido and Y. Tokura, "Insulator-metal transition and giant magnetoresistance in $\mathrm{La}_{1-x}\mathrm{Sr}_x\mathrm{MnO}_3$," *Phys. Rev. B,* vol. 51, no. 20, p. 14103–14109, May 1995.

[37] H. Fujiwara, Spectroscopic Ellipsometry: Principles and Applications, John Wiley &Sons, 2003.